\documentclass[11pt]{article}

\usepackage{geometry}
\usepackage[utf8]{inputenc}
\usepackage{cloud_preamble}
\usepackage{enumitem}
\usepackage[ruled,vlined]{algorithm2e}
\usepackage{booktabs}
\usepackage{tikz}

\usepackage{draftwatermark}
\SetWatermarkText{DRAFT}
\SetWatermarkScale{5}
\SetWatermarkLightness{0.95}

\usetikzlibrary{arrows}
\usetikzlibrary{calc}

\usepackage[sorting=ynt]{biblatex} 
\newcommand{\player}{\text{P}^{\text{-}}} 

\newcommand{\actind}{I}
\newcommand{\allplayers}{{\mathcal{N}\cup\{\text{c}\}}} 
\newcommand{\allactions}{\mathbf{A}} 

\title{Variance decompositions for extensive-form games}
\author{Alex Cloud, Eric Laber\\ North Carolina State University}
\date{September 8, 2020}

\numberwithin{equation}{section} 
\begin{document}

\maketitle

\begin{abstract}
    Quantitative measures of randomness in games are useful for game design and have implications for gambling law. We treat the outcome of a game as a random variable and derive a closed-form expression and estimator for the variance in the outcome attributable to a player of the game. We analyze poker hands to show that randomness in the cards dealt has little influence on the outcomes of each hand. A simple example is given to demonstrate how variance decompositions can be used to measure other interesting properties of games.
\end{abstract}

\section{Introduction}
From game design studios to courtrooms, randomness in games has been the subject of extensive discussion. Game designers use random game elements to protect players' egos, increase gameplay variety, and limit the efficacy of mental calculation \cite{elias2012characteristics}. In U.S. state law, the question of whether Poker is predominantly a game of chance or skill is considered to be central to the legality of online Poker \cite{kelly2007poker,levitt2014role}. 

The question of how to measure the role of luck versus skill has proved difficult
and produced many answers \cite{dedonno2008poker,croson2008poker,elias2012characteristics, levitt2014role,
heubeck2008measuring, heubeck2008measuring2, getty2018luck}. 
For example, in {\em USA v. Lawrence Dicristina}, economic consultant and high-level amateur poker player Randal Heeb testified 
that ``statistical analysis of poker hands confirms that skill predominates over chance.'' His conclusion was based on a series of
heuristic data analyses combined with intuitive judgments \cite{heeb2012report}. Others have argued that the strong association 
between player skill rating and future earnings constitute strong evidence that poker should be considered a game of skill 
\cite{levitt2014role,van2015beyond}. 

A first step in assessing the role of chance in a game is to quantify sources of uncertainty. We examine how variation in the outcomes of a game can be attributed to players or chance events using a variance decomposition, a standard statistical method in which the variance of a random variable is written as the sum of nonnegative terms corresponding to variation attributable to different factors \cite{fisher1919xv}. We express the total variation in game outcomes as the sum of 
variance components associated with (i) the actions taken by a player of interest, and (ii) all remaining sources of variation. 
By applying this decomposition to a conceptual ``chance player,'' we measure the degree to which randomness inherent in a game biases the results in favor of a given player. 

We derive an analytical expression for the variance components and use it to obtain estimators which are model-free in the sense that they do not require access to an entire game model or other players' behavior. Our results apply to finite extensive-form games in general; they are not limited to the two-player, zero-sum case. As an illustrative example, we analyze poker hands played by the DeepStack poker AI against professional players \cite{moravvcik2017deepstack} and find that chance events have very little influence on the expected per-hand profit for a player relative to the total variation in per-hand profit.  

\section{Extensive-form games}
An {\em extensive-form game} is a tree-based representation of a multi-agent system; Figure \ref{fig:extensive_game}
displays a simple example.
In this representation, the game is played by traversing the tree from the root to a leaf node, 
with a player's action at each node determining the next node visited. Our notation is based on \cite{lanctot2009monte} and \cite{heinrich2015fictitious}, with some modifications.

\begin{figure} 
\centering
\begin{tikzpicture}
\tikzstyle{hollow node}=[circle,draw,inner sep=1.5]
\tikzstyle{solid node}=[circle,draw,inner sep=1.5,fill=black]
\tikzstyle{level 1}=[level distance=10mm,sibling distance=3cm]
\tikzstyle{level 2}=[level distance=10mm,sibling distance=1.5cm]
\tikzstyle{level 3}=[level distance=10mm,sibling distance=1cm]

\node[solid node, label=above right:{c}]{}
child{
    node(1)[solid node, label= above left:{1}]{}
    child{
        node(11)[solid node, label=above left:{2}]{} 
        child{
            node[hollow node,label=below:{$0$}]{}
        }
        child{
            node[hollow node,label=below:{$-1$}]{}
        }
    }
    child{
        node(12)[solid node]{} 
        child{
            node[hollow node,label=below:{$1$}]{}
        }
        child{
            node[hollow node,label=below:{$0$}]{}
        }
    }
    edge from parent[] node[yshift=7,xshift=-7]{0.5}
}
child{
    node(2)[solid node, label=above right:{c}]{}
    child{node[hollow node,label=below:{$1$}]{} edge from parent[] node[yshift=2,xshift=-11]{0.5} }
    child{node[hollow node,label=below:{$-1$}]{} edge from parent[] node[yshift=2,xshift=11]{0.5} }
    edge from parent[] node[yshift=7,xshift=7]{0.5}
}
;

\draw[dashed,rounded corners=8]($(11) + (-.25,.25)$) rectangle ($(12) +(.25,-.25)$);

\end{tikzpicture}

\caption{An example of an extensive-form game. Each node in the tree is a state $s \in \St$ and is annotated with the corresponding player, $\player(s)$. The dashed line represents Player 2's information state; in this example, they cannot tell what move Player 1 played. Rewards for Player 1 are shown below the terminal nodes.} \label{fig:extensive_game}
\end{figure}

Let $\St$ denote the set of possible game states which we assume is finite; each state is associated
with a node in the game tree.  
Define
$\mathcal{N} = \left\lbrace 1,\ldots, n\right\rbrace$ to be the set of (non-chance) players and
let $c$ denote the chance player.  
The {\em player function} $\player: \mathcal{S} \rightarrow \allplayers$ associates each state with a player\textbf{}. At each state $s \in \St$, there are a finite number of available actions $\A(s)$, such that each $a \in \A(s)$ uniquely determines the next state visited in the tree \cite{shoham2008multiagent}.

A sequence of actions $z=(a_1,\dots,a_m)$ is a {\em terminal history} if it leads from the root to a leaf of the game tree; let $\mathcal{Z}$ denote the set of all terminal histories. For each player $i \in \mathcal{N}$ and  terminal history $z \in \mathcal{Z}$, a reward $r^i(z) \in \R$ is obtained by player $i$ upon reaching $z$. Each player $i \in \mathcal{N}$ has a set of {\em information states} $\,\U^i$ which represent collections of nodes which are indistinguishable to the player. In particular, $\U^i$ is a partition of $\{ s \in \St : \player(s) = i\}$ with the additional condition that $\A(s) = \A(s')$ if $s$ and $s'$ are in the same information state. So, we can write $\A(u)$ for $u \in \U^i$ unambiguously. Define $\U^{\text{c}} = \{ \{s \} \, : \, \player(s) = \text{c} \}$. We consider games of {\em perfect recall}, so that for every player $i$, each $u \in \U^i$ can be uniquely identified with the a sequence of information states and actions required to arrive there. 

Finally, the behavior of each player $i \in \allplayers$ is described by a {\em policy} $\pi^i$ (also known as a behavioral 
strategy), which is a function that maps each information state $u \in \U^i$ to a distribution over the allowable 
actions $\A(u)$. A {\em policy profile} is a tuple of player policies, $\pi = (\pi^1,\dots,\pi^n)$. By convention, the policy of 
the chance player $\pi^\text{c}$ is considered to be a fixed part of the extensive-form game itself and not a part of any policy 
profile. 

\section{Variance decompositions for game outcomes} \label{sec:var_decomp}

\subsection{Extensive-form games with random variables}
For convenience, we introduce random variables that represent 
the actions selected by players in a single play of the game. For each $i \in \allplayers$, 
and for each $u \in \U^i$, let $A(u)$ be a random variable taking values in $\A(u)$ which represents the action 
player $i$ would take given information state $u$. 
This variable always realizes a value, even if $u$ is not reached in a particular 
play of the game. Note that for $u \neq u' \in \U^i$, it need not be the case that $A(u)$ is independent of $A(u')$. This way 
specifying player behavior is quite general and can account for different models of player action selection. For example, a player
may randomly precommit to a deterministic policy (this is known as a mixed strategy in the game theory literature), or select 
actions independently at random at each time step (a behavioral strategy) \cite{koller1996finding}.

For each terminal history $z \in \mathcal{Z}$ and player $i \in \allplayers$, let $m^i(z)$ be the number of actions selected by player $i$ along $z$, so that for each $j \in \{1, \dots, m^i(z)\}$, we can write $u^i_{z,j}$ and $a^i_{z,j}$ to denote the $j$th information state observed and action selected by player $i$ along terminal history $z$. Define $\actind^i_{z,j} = \ind[A(u_{z,j}^i) = a^i_{z,j}]$ to be the Bernoulli random variable that indicates whether player $i$ selects $a^i_{z,j}$ at $u^i_{z,j}$. Finally, define $\actind_{z}^i = \prod_{j=1}^{m^i(z)} \actind^i_{z,j}$ to be the Bernoulli random variable that indicates whether player $i$ selects all actions along $z$. (If $m^i(z) = 0$, set $\actind_{z}^i \equiv 1$.)

A terminal history occurs if and only if every action along it is selected. Therefore, for each $z \in \mathcal{Z}$, $I_z = \prod_{i \in \allplayers} \actind^i_z $ defines a Bernoulli random variable such that the success probability $P(I_z = 1)$ is the probability that terminal history $z$ is realized. Let $Z$ be a random terminal history variable such that $P(Z=z) = P(I_z = 1)$ for all $z$ that represents a random play-through of the game. This allows us to cast the outcome of an extensive-form game as 
\begin{align*}
\mathbf{Y} = [r^1(Z),\dots,r^n(Z)].
\end{align*}
Write $Y = r(Z) = r^i(Z)$, the random reward for a particular player of interest $i \in \allplayers$ upon a play of the game. Our goal is to express its variance, $V(Y)=E\{[Y-E(Y)]^2\}$, as a sum of nonnegative terms corresponding to meaningful properties of a game.

\subsection{Variance decomposition} \label{sec:variance_decomposition_results}
Let $i \in \allplayers$ be a player of interest, and let $\allactions^i = [A(u)]_{u \in \U^i}$ be the concatenation of all actions for player $i$. By the law of total variance we can decompose the variance in game outcomes as
\begin{align}
    V(Y) = V[E(Y|\allactions^i)] + E[V(Y|\allactions^i)]. \label{eqn:law_of_total_variance}
\end{align}
The term $E(Y|\allactions^i)$ is the average game outcome upon many traversals of the game tree when player $i$ commits ahead of time to playing the actions in $\allactions^i$. For example, $E(Y|\allactions^\text{c})$ represents the average outcome for a group of poker players who play the same hand from a deck with a particular card order many times, or the average outcome for a pair of chess players who start with the same colors every game. Then $V[E(Y|\allactions^\text{c})]$ is the variation in this mean as the chance actions $\allactions^\text{c}$ vary, and represents the variation in game outcomes due to chance events. The latter term of \eqref{eqn:law_of_total_variance} has a similar interpretation as the variation in game outcomes not explained by actions selected by player $i$.

Let $i \in \allplayers$ be a player of interest. Suppose that player $i$ plays according to a behavioral strategy $\pi^i$, meaning that $A(u)$ is independent of $A(u')$ for all $u \neq u' \in \U^i$ and action probabilities are given by a policy such that $P(A^i_{z,k} = 1) = \pi^i(a^i_{z,k} | u^i_{z,k})$ for all $z \in \mathcal{Z}$ and $k \in \{1,\dots,m^i(z)\}$. No such assumption is required for the remaining players; we only require that their actions be independent of the actions of player $i$. 

For $i \in \allplayers$, define $\eta^i(z) = P(\actind^i_z = 1)$,  $\eta^{-i}(z) = P(\prod_{i' \in \allplayers \setminus \{i\}} \actind^{i'}_z = 1)$, and $\eta(z) = \eta^i(z) \, \eta^{i}(z) = P(I_z = 1)$. For each information state $u \in \U^i$, define $\mathcal{Z}(u) = \{ z \in \mathcal{Z} : u \text{ is visited in } z\}$ and for each $a \in \A(u)$, define $\mathcal{Z}(ua) = \{ z \in \mathcal{Z} : u \text{ is visited in } z \text{ and action } a \text{ is selected at } u.\}$. Define $q(u,a) = E[r(Z) | Z \in \mathcal{Z}(u,a)]$ to be the expected outcome given that player $i$ is at $u$ and takes action $a$; similarly, define $v(u) = E[r(Z) | Z \in \mathcal{Z}(u)]$. 

Our main result is an expression of the variance in game outcomes explained by player $i$'s actions as a sum of weighted, squared action-value and value functions over all of player $i$'s information states:
\begin{align}
     V[E(Y|\mathbf{A}^i)] = \sum_{u \in \U^i}  \bigg( \sum_{a \in \A(u)} [q(u,a) ]^2 \, \pi^i(a | u) - [v(u)]^2 \bigg) \, \eta^{-i}(u) \, \eta(u). \label{eqn:info_state_based_component}
\end{align}
A proof is provided in Appendix \ref{sec:decomp_derivation_value_based}. Computing this requires traversing the game tree a fixed number of times and hence is $O(|\St|)$. From this we obtain a formula for the other variance component by observing that $E[V(Y|\allactions^i)] = V(Y) -  V[E(Y|\mathbf{A}^i)]$, where $V(Y)$ can be evaluated as $\sum_{z \in \mathcal{Z}} [ r(z) - \sum_{z' \in \mathcal{Z}} r(z') \eta(z')]^2 \, \eta(z)$.

Assuming $\eta^{-i}$, $q$, and $v$ are known, given an i.i.d. sequence of $\nu$ playthroughs of the game, each generating a sequence $\overline{U}_k = (U_{k,1},\dots,U_{k,l_k})$ of observed information states in $\U^i$, then the following is a consistent estimator for $V[E(Y|\mathbf{A}^i)]$ as proved in Appendix \ref{sec:consistency_proof}:
\begin{align}
   \nu^{-1} \sum_{k=1}^\nu \sum_{l=1}^{l_k} \Big( \sum_{a \in \A(U_{k,l})} [q(U_{k,l},a) ]^2 \pi^i(a|U_{k,l}) - [v(U_{k,l})]^2 \Big) \, \eta^{-i}(U_{k,l}).
\end{align}
In practice, $q$ and $v$ can be estimated by supervised learning and $\eta^i = \eta / \eta^{-i}$ can be estimated with $\est{\eta}(u) = \nu^{-1}\sum_{k=1}^\nu \sum_{l=1}^{l_k} \ind(U_{k,l} = u)$ and $\eta^i(u) = \pi^i(u)$ (assuming the analyst does not have access to opponent policies and observations). However, if there are many possible information states, i.e. $|\U^i|$ is large, $\est{\eta}(u)$ will greatly overestimate the visit probability. An alternative is the more straightforward regression-based estimator. The regression-based estimator works by fitting a model for the conditional mean of the game outcome given a player's actions, then computing the empirical variance of the conditional mean estimator. The procedure is:
\begin{enumerate}
    \item Specify a parametric model $f_\theta$ that maps the collection of all actions for the player of interest to a real number, $f_\theta : \times_{u \in \U^i} \A(u) \rightarrow \R$.
    \item For a each observed game $k \in \{ 1,\dots,\nu \}$, record action-outcome pairs $(\allactions^i_k, Y_k)$. For each $k$, if an information state for the player of interest, $u \in \U^i$ was not visited in game $k$, sample $A(u) \sim \pi^i(\cdot|u)$ and include the sampled action in $\allactions^i_k$.
    \item Fit the model on the action-outcome pair data to find a $\est{\theta}$ that minimizes the mean square error, $\nu^{-1} \sum_{k=1}^{\nu} [ Y_k - f_{\widehat{\theta}}(\allactions^i_k)]^2$, so $f_{\widehat{\theta}}(\cdot)$ estimates $E(Y|\mathbf{A}^i=\cdot)$.
    \item Compute the empirical variance of $f_{\widehat{\theta}}(\allactions^i)$, which is $\nu^{-1} \sum_{k=1}^{\nu} [ f_{\widehat{\theta}}(\allactions^i_k) - \nu^{-1} \sum_{h=1}^\nu f_{\widehat{\theta}}(\allactions^i_h) ]^2$. This is our estimate of $V[E(Y|\mathbf{A}^i)]$.
\end{enumerate}

\section{Analysis of professional poker players versus DeepStack}

We analyze 150 thousand hands of heads-up no-limit poker played by different players against the Poker AI Deepstack, including 45 thousand hands played by self-identified professional players. For details on how the data were generated, see the supplemental materials of the DeepStack paper \cite{moravvcik2017deepstack}. Our goal is to understand the role chance has in influencing the per-hand profits of a human playing against Deepstack, so we will estimate the variance component for chance for games played by each human player indexed by $j \in \{1, \dots, 33\}$. We also include an algorithm used for poker agent evaluation called Local Best Response (index $j=0$), which we include as a form of transfer learning in order to improve estimates of expected outcomes for the human players. Assume that player $j$ plays according to a policy $\pi_j$ and write $E_{\pi_j}(Y|\allactions^{\text{c}})$ to denote the expected per-hand profit for player $j$ against DeepStack given all chance events $\allactions^{\text{c}}$. Then we would like to know $V[E_{\pi_j}(Y|\allactions^{\text{c}})]$ for each $j$.

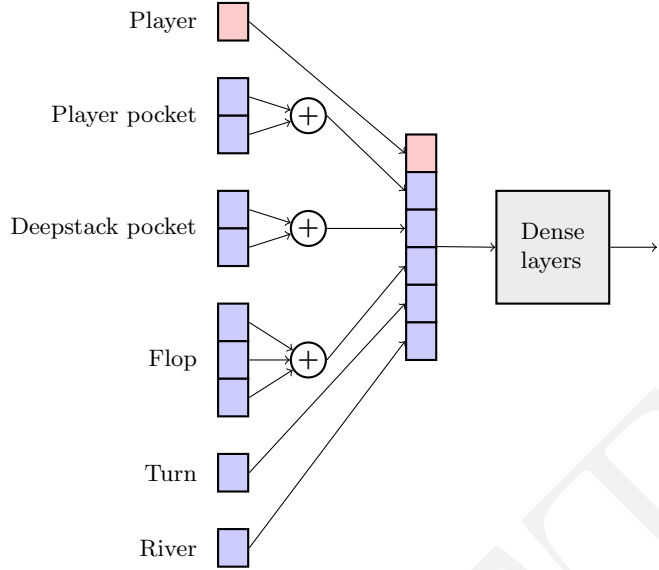
\begin{figure}
    \centering
    \footnotesize
    \begin{tikzpicture}[
    roundnode/.style={circle, draw=black, thick, minimum size=3mm, inner sep =-3mm},
    playernode/.style={rectangle , draw=black, fill=red!20, thick, minimum width=4mm, minimum height=5mm},
    cardnode/.style={rectangle, draw=black, fill=blue!20, thick, minimum width=4mm, minimum height=5mm},
    densenode/.style={rectangle, draw=black, fill=gray!15, thick, minimum width=1.5cm, minimum height=1.5cm}
    ]
    \node[label=left:Player] at (-0.25, 7) (playerlabel) {};
    \node[label=left:Player pocket] at (-0.25, 5.75) (playerpocketlabel) {};
    \node[label=left:Deepstack pocket] at (-0.25, 4.25) (deepstackpocketlabel) {};
    \node[label=left:Flop] at (-0.25, 2.5)(floplabel) {};
    \node[label=left:Turn] at (-0.25, 1) (turnlabel) {};
    \node[label=left:River] at (-0.25, 0) (riverlabel) {};
    
    \node[playernode] at (0,7)   (player) {};
    
    \node[cardnode]   at (0,6)   (playerpocket1) {};
    \node[cardnode]   at (0,5.5) (playerpocket2) {};
    
    \node[cardnode]   at (0,4.5) (deepstackpocket1) {};
    \node[cardnode]   at (0,4)   (deepstackpocket2) {};
    
    \node[cardnode]   at (0,3)   (flop1) {};
    \node[cardnode]   at (0,2.5) (flop2) {};
    \node[cardnode]   at (0,2)   (flop3) {};
    
    \node[cardnode]   at (0,1)   (turn) {};

    \node[cardnode]   at (0,0)   (river) {};
    
    \node[playernode] at (2.5, 5.25) (playerconcat) {};
    \node[cardnode] at (2.5, 4.75) (playerpocketconcat) {};
    \node[cardnode] at (2.5, 4.25) (deepstackpocketconcat) {};
    \node[cardnode] at (2.5, 3.75) (flopconcat) {};
    \node[cardnode] at (2.5, 3.25) (turnconcat) {};
    \node[cardnode] at (2.5, 2.75) (riverconcat) {};
    
    \node[roundnode] at (1, 5.75) (sumplayerpocket) {\bf +};
    \node[roundnode] at (1, 4.25) (sumdeepstackpocket) {\bf +};
    \node[roundnode] at (1, 2.5)  (sumflop) {\bf +};

    \draw[->] (playerpocket1.east) -- (sumplayerpocket);
    \draw[->] (playerpocket2.east) -- (sumplayerpocket);
    
    \draw[->] (deepstackpocket1.east) -- (sumdeepstackpocket);
    \draw[->] (deepstackpocket2.east) -- (sumdeepstackpocket);
    
    \draw[->] (flop1.east) -- (sumflop);
    \draw[->] (flop2.east) -- (sumflop);
    \draw[->] (flop3.east) -- (sumflop);
    
    \draw[->] (player.east) -- (playerconcat.west);
    \draw[->] (sumplayerpocket.east) -- (playerpocketconcat.west);
    \draw[->] (sumdeepstackpocket.east) -- (deepstackpocketconcat.west);
    \draw[->] (sumflop.east) -- (flopconcat.west);
    \draw[->] (turn.east) -- (turnconcat.west);
    \draw[->] (river.east) -- (riverconcat.west);
    
    \node[densenode, align=left] at (4.25, 4) (denselayers) {Dense \\layers};
    
    \node at (2.5, 4.01) (concatcenter) {};
    \draw[->] (concatcenter.east) -- (denselayers.west);
    
    \node at (5.75, 4) (output) {};
    \draw[->] (denselayers.east) -- (output.west);
    
    \end{tikzpicture}
    \caption{The neural network architecture used for analysis of Deepstack hands. The input for each card (shown in blue) is a concatenation of the rank and suit of the card. The rank and suit are each assigned a vector embedding, with the same weights shared for all card inputs.}
    \label{fig:deepstack_analysis_architecture}
\end{figure}

We use a neural network to estimate $E_{\pi_j}(Y|\allactions^{\text{c}})$ given a player and the realization of all chance events:
\begin{itemize} \itemsep0em
    \item The player's pocket cards (2 cards)
    \item Deepstack's pocket cards (2 cards)
    \item The flop (3 cards)
    \item The turn (1 card)
    \item The river (1 card)
\end{itemize}
The neural network shares a representation of cards across all inputs: each card rank (e.g. Ace) and suit (e.g. hearts) is associated with a learned vector embedding; a card is represented by the concatenation of these embeddings. To capture the unordered nature of players' pocket cards and the flop, the card representations for each of those groups is summed. The architecture is depicted in Figure \ref{fig:deepstack_analysis_architecture}.

Our model was trained by stochastic gradient descent with the Adam optimizer \cite{kingma2014adam} with early stopping based on cross-validation loss using a 90\%/10\% train-test split. If a hand ended before all chance events were observed (for example, if a player folded before the river), the cards associated with that chance event were randomly sampled from the remaining cards in the deck at that point in the game. These cards were resampled in each epoch of training in order to decrease variance. We present our results in Table \ref{tab:deepstack}. For each player, the empirical variance of the regression estimator was computed over both the training and test data and is recorded in column ``Chance var.''

Typical values for the percent of total variance ``explained'' by chance events fall between 0\% and 2\%. We conclude that the influence of chance events alone on per-hand outcomes is quite limited. Rather, the large amount of variation in per-hand profits is mostly explained by player randomization and the interaction between those actions and chance. We elaborate in the Discussion section. 
\begin{table}
\footnotesize
\centering
\begin{tabular}{lrrrrr}
\toprule
          Player name &  Num. hands &  Mean profit &       Variance. &  Chance var. &  Chance var. \% \\
\midrule
  Local best response &    106221 &        -66.2 &  4711362.4 &      68577.4 &            1.5 \\
        Ivan Shabalin &      3122 &        -33.5 &  3419341.6 &      26105.1 &            0.8 \\
             Pol Dmit &      3026 &        -93.3 &  4992102.7 &      78447.3 &            1.6 \\
         Muskan Sethi &      3010 &       -214.1 &  8069582.3 &     152633.7 &            1.9 \\
        Dmitry Lesnoy &      3007 &         11.5 &  4422563.8 &      27593.1 &            0.6 \\
   Stanislav Voloshin &      3006 &          6.4 &  3272512.8 &      19559.7 &            0.6 \\
      Lucas Schaumann &      3004 &        -15.7 &  2585189.5 &      38743.3 &            1.5 \\
            Phil Laak &      3003 &        -77.3 &  3576155.5 &      37342.1 &            1.0 \\
 Antonio Parlavecchio &      3003 &       -108.8 &  7218296.0 &     116614.0 &            1.6 \\
           Kaishi Sun &      3002 &         -0.5 &  4138381.9 &      36490.0 &            0.9 \\
         Martin Sturc &      3001 &         51.3 &  2579614.6 &      26995.9 &            1.0 \\
  Prakshat Shrimankar &      3001 &        -17.4 &  3468418.3 &      43344.5 &            1.2 \\
      Tsuneaki Takeda &      1901 &         33.3 &  7458278.7 &      13850.9 &            0.2 \\
           Youwei Qin &      1759 &       -195.3 & 14797348.3 &     118693.4 &            0.8 \\
         Fintan Gavin &      1555 &          2.6 & 10967917.3 &      35274.5 &            0.3 \\
     Giedrius Talacka &      1514 &        -45.9 & 11464541.4 &      69281.2 &            0.6 \\
     Juergen Bachmann &      1088 &       -176.9 &  7804660.5 &     190849.5 &            2.4 \\
       Sergey Indenok &       852 &        -25.3 & 13895176.8 &      44966.3 &            0.3 \\
     Sebastian Schwab &       516 &       -180.0 &  6250606.2 &      25266.9 &            0.4 \\
        Dara Okearney &       456 &        -22.3 &  3365433.0 &      30225.4 &            0.9 \\
   Roman Shaposhnikov &       330 &         89.8 &  3951695.5 &      22054.5 &            0.6 \\
            Shai Zurr &       330 &       -115.4 &  4148165.0 &      43651.8 &            1.1 \\
       Luca Moschitta &       328 &       -143.8 &  4833549.3 &      76582.4 &            1.6 \\
      Stas Tishekvich &       295 &         34.6 &  3904856.9 &      68538.0 &            1.8 \\
          Eyal Eshkar &       191 &        -71.5 &  8773209.3 &      83846.6 &            1.0 \\
          Jefri Islam &       176 &       -382.2 & 10558538.8 &      67446.8 &            0.6 \\
              Fan Sun &       122 &        129.1 &  9265866.5 &      54513.5 &            0.6 \\
        Igor Naumenko &       102 &        -85.1 &   611611.4 &      32192.0 &            5.3 \\
    Silvio Pizzarello &        90 &       -513.4 & 10435348.4 &      96080.5 &            0.9 \\
          Gaia Freire &        76 &        -13.8 &    92173.2 &      44686.1 &           48.5 \\
        Alexander Bös &        74 &         -0.1 &  1286240.1 &       7153.1 &            0.6 \\
        Victor Santos &        58 &        175.9 &   956344.0 &      24456.7 &            2.6 \\
            Mike Phan &        32 &       1122.2 & 25579723.6 &      25381.5 &            0.1 \\
   Juan-Manuel Pastor &         7 &       -728.6 &  1135714.3 &      26707.9 &            2.4 \\
\bottomrule
\end{tabular}
\caption{Analysis of expected per-hand player profits for human professionals against the DeepStack Poker AI.} \label{tab:deepstack}
\end{table}

\section{A three-way decomposition for assessing skillfulness of a game}
As another example of using variance decompositions to analyze games, we present a concept for measuring skill, chance, and non-transitivity that is inspired by prior work on decompositions of games \cite{candogan2011flows} and recent developments regarding learning in the context of complex games with nontransitive elements \cite{balduzzi2019open} \cite{omidshafiei2020navigating}. For simplicity, assume we are given a symmetric two-player zero-sum game and a population of players represented by a finite set of policies $\Pi$, each with a skill rating $\rho_\pi$ for $\pi \in \Pi$. One notion of the skillfulness of the game is the variance in outcomes explained by players' skill ratings alone, assuming two policies $(\pi_1,\pi_2)$ are sampled uniformly from $\Pi$:
\begin{align}
    V(Y) = V[E(Y|\rho_{\pi_1},\rho_{\pi_2})] + E[V(Y|\rho_{\pi_1},\rho_{\pi_2})]. \label{eqn:skill_decomp}
\end{align}
\sloppy Applying the law of total variance to the conditional variance $V(Y|\rho_{\pi_1},\rho_{\pi_2})$, we condition on chance actions $\allactions^\text{c}$ as in \eqref{eqn:law_of_total_variance} to obtain $V(Y|\rho_{\pi_1},\rho_{\pi_2}) = V[E(Y|\allactions^\text{c},\rho_{\pi_1},\rho_{\pi_2})|\rho_{\pi_1},\rho_{\pi_2}] +
E[V(Y|\allactions^\text{c},\rho_{\pi_1},\rho_{\pi_2})|\rho_{\pi_1},\rho_{\pi_2}]$. Using linearity of expectation and the tower rule, this allows us to extend \eqref{eqn:skill_decomp} to
\begin{align}
        V(Y) = \underbrace{V[E(Y|\rho_{\pi_1},\rho_{\pi_2})]}_{\text{skill}}+ \underbrace{E\{V[E(Y|\allactions^\text{c},\rho_{\pi_1},\rho_{\pi_2})|\rho_{\pi_1},\rho_{\pi_2}]\}}_{\text{chance}}+ \underbrace{E[V(Y|\allactions^\text{c},\rho_{\pi_1},\rho_{\pi_2})]}_{\text{remaining variation}}. \label{eqn:threeway_decomp}
\end{align}

We apply this formula to analyze a simple game parametrized by constants $n \in \N$, $c \in \N \cup \{0\}$, and $\alpha \in [0,1]$ that can be seen as an abstract model of a game with a skill component (some strategies are strictly better than others), a nontransitive component (there exist cycles of pure strategies), and chance (some games are decided by events entirely out of the players' hands). Skillful Rock Paper Scissors, or SkillRPS($n$, $c$, $\alpha$), is defined as follows: each player $i \in \{1,2\}$ simultaneously selects a number $N_i \in \{1,\dots, n\}$ and a move $A_i \in \{ \text{Rock, Paper, Scissors} \}$. Player 1's score is $S=N_1-N_2+c \cdot \text{RPS}(A_1,A_2)$, where RPS is the payoff function for Rock Paper Scissors depicted in Table \ref{tab:rps}.

\begin{table}
\centering
\begin{tabular}{r|ccc}
 $a_1$\textbackslash $a_2$ & Rock & Paper & Scissors \\ \hline
Rock     & 0    & -1    & 1 \\
Paper    & 1    & 0     & -1 \\
Scissors & -1   & 1     & 0 
\end{tabular}
\caption{The payoff function RPS($a_1$,$a_2$).} \label{tab:rps}
\end{table}

The outcome of the game for player 1 is $Y= (1-W)[\ind(S>0)-\ind(S<0)] + W(2Z-1)$, where $W \sim \text{Bernoulli}(\alpha)$ and $Z \sim \text{Bernoulli}(1/2)$ are chance events such that $W$ determines whether the game is decided by a fair coin flip $Z$. Note that when $n=1$, $c>0$, $\alpha=0$, the game is classic Rock Paper scissors, when $\alpha=1$ it is a coin flip, and when $c=0$ it is a transitive game. 
The game can be represented in extensive form as shown in Figure \ref{fig:extensive_game}, which depicts SkillRPS(2, 0, $0.5$).

\begin{figure}
    \centering
    \hbox{\hspace{0.4cm} \includegraphics{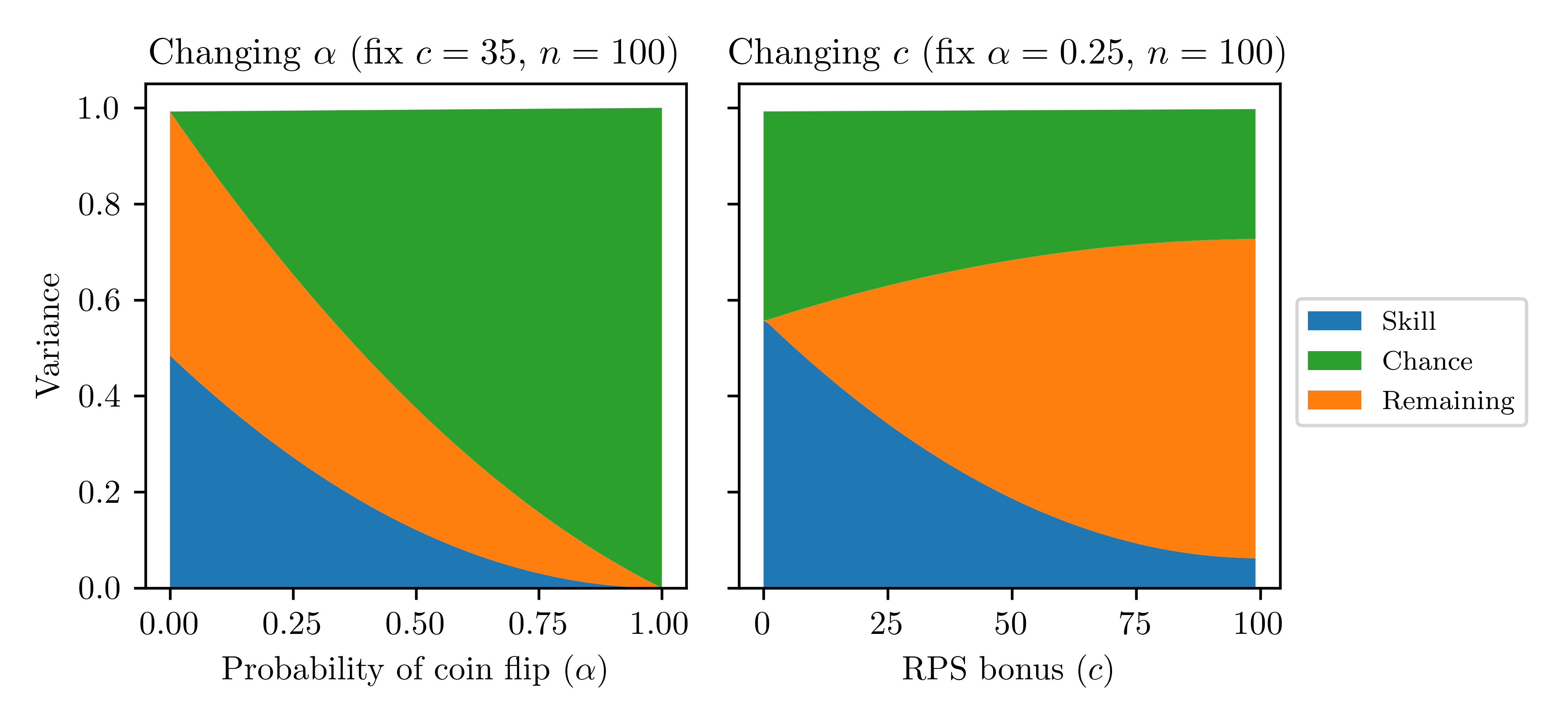}}
    \caption{Three-way variance decompositions for SkillRPS with different game parameters under the assumption that players selects moves independently and uniformly at random, i.e. for $i \in \{1,2\}$, $N_i \sim \text{Uniform}(\{1,\dots,n\})$ and $A_i \sim \text{Uniform}(\{ \text{Rock, Paper, Scissors}\})$ and are independent. Details on the variance components for SkillRPS are included in Appendix \ref{sec:skillrps_details}.  }
    \label{fig:skillrps_decomp}
\end{figure}

In Figure \ref{fig:skillrps_decomp}, the three-way decomposition is given across many values of the SkillRPS game parameters, showing that the components correspond to meaningful properties of games: increasing the probability that the game outcome is determined by a coin flip increases the chance variance component to 1 as the other variance components decrease smoothly; increasing the bonus for winning at Rock Paper Scissors decreases the skill component. In this case, the ``remaining'' variation corresponds directly to the non-transitivity introduced by the RPS component of the game. 

\section{Discussion}


One might hope that the variance component for chance $V[E(Y|\allactions^\text{c})]$ measures how lucky a game is in the context of the players playing the game. We argue that this is not the case, and conclude with thoughts on the applicability of variance component estimation for the analysis of games.

First, the variance component for chance does not measure how lucky a game is because by design it avoids measuring variation introduced by random player actions. Consider the classic version of Rock Paper Scissors (RPS) depicted in Table \ref{tab:rps}. A cautious player can guarantee an expected payoff of 0 by assigning uniform probability to each action, causing the outcome of the game to be uniformly random over $\{-1,0,1\}$. For this reason, it is natural to view RPS as a game of luck--- however, RPS as typically modeled does not have a chance player. All variation in RPS comes from randomness in player action selection. So, if we are to call RPS a game of luck, then a notion of luck that only considers chance events is inadequate.

Second, the variance component for chance is conservative in that it only measures the marginal (average) effect of chance actions on game outcomes. It does not capture the interaction between chance events and player actions. For example, consider a variant of RPS in which one of the players is replaced with a chance player. If the non-chance player employs a uniform random policy, then the expected outcome is 0 regardless of action is selected by chance. Thus $E(Y | \allactions^\text{c} = a) = 0$ for each $a \in \{ \text{Rock}, \text{Paper}, \text{Scissors} \}$. This means that for any chance policy, the variance component for chance is 0, yet from the player's perspective, against a uniform chance policy, it is as though the game outcome is entirely determined by chance! 

What the variance component for chance actually measures is the per-game amount that chance biases the outcome in favor of a player. In both the examples given above, luck plays a significant role in the game outcomes, but the realization of chance events alone does not tend to significantly tilt the game in the favor of either player-- so our measure evaluates to 0. Returning to the analysis of DeepStack poker hands, we can see that despite the large amount of variation in per-hand profits (of which any one realization could be called ``lucky'') the game (as played at a high level) is in some sense fair: on a hand-by-hand basis, the average amount that the random deck order advantages or disadvantages a particular player is small.

Video game designers may find the variance component for chance helpful in assessing the per-play advantage gleaned by a player due to chance events. We speculate that for a rewarding game experience, the variance component should be kept low, or else players will a sense of limited agency. Returning to the question of the legality of poker, our measure could represent a sufficient (but not necessary) criterion for determining that a game is ``predominantly due to chance:" if the ratio of the variance component for the chance player to the total variation is greater than 50\%, then clearly the game outcomes could be said to be predominantly due to chance. The three-way variance decomposition in \eqref{eqn:threeway_decomp} offers a way to characterize meaningful properties of games that arise in the context of multiagent reinforcement learning and presents new research challenges such as (i) accounting for estimation error in the skill rating (however it is defined), and (ii) accounting for the actual distribution from which policies are sampled to play each other, which is often not uniform but rather skill-based, such that players with nearby skill ratings are likely to be placed together.

 
\printbibliography

@article{levitt2014role,
  title={The role of skill versus luck in poker evidence from the world series of poker},
  author={Levitt, Steven D and Miles, Thomas J},
  journal={Journal of Sports Economics},
  volume={15},
  number={1},
  pages={31--44},
  year={2014},
  publisher={Sage Publications Sage CA: Los Angeles, CA}
}

@article{heubeck2008measuring,
  title={Measuring skill in games: A critical review of methodologies},
  author={Heubeck, Steven},
  journal={Gaming Law Review and Economics},
  volume={12},
  number={3},
  pages={231--238},
  year={2008},
  publisher={Mary Ann Liebert, Inc. 140 Huguenot Street, 3rd Floor New Rochelle, NY 10801~…}
}

@article{heubeck2008measuring2,
  title={Measuring skill in games with random payoffs: Evaluating legality},
  author={Heubeck, Steven},
  journal={Review of Law \& Economics},
  volume={4},
  number={1},
  pages={25--34},
  year={2008},
  publisher={De Gruyter}
}

@article{getty2018luck,
  title={Luck and the law: Quantifying chance in fantasy sports and other contests},
  author={Getty, Daniel and Li, Hao and Yano, Masayuki and Gao, Charles and Hosoi, AE},
  journal={SIAM Review},
  volume={60},
  number={4},
  pages={869--887},
  year={2018},
  publisher={SIAM}
}

@article{fisher1919xv,
  title={XV.—The correlation between relatives on the supposition of Mendelian inheritance.},
  author={Fisher, Ronald A},
  journal={Earth and Environmental Science Transactions of the Royal Society of Edinburgh},
  volume={52},
  number={2},
  pages={399--433},
  year={1919},
  publisher={Royal Society of Edinburgh Scotland Foundation}
}

@book{shoham2008multiagent,
  title={Multiagent systems: Algorithmic, game-theoretic, and logical foundations},
  author={Shoham, Yoav and Leyton-Brown, Kevin},
  year={2008},
  publisher={Cambridge University Press}
}

@article{kingma2014adam,
  title={Adam: A method for stochastic optimization},
  author={Kingma, Diederik P and Ba, Jimmy},
  journal={arXiv preprint arXiv:1412.6980},
  year={2014}
}

@article{moravvcik2017deepstack,
  title={Deepstack: Expert-level artificial intelligence in heads-up no-limit poker},
  author={Moravcik, Matej and Schmid, Martin and Burch, Neil and Lis{\`y}, Viliam and Morrill, Dustin and Bard, Nolan and Davis, Trevor and Waugh, Kevin and Johanson, Michael and Bowling, Michael},
  journal={Science},
  volume={356},
  number={6337},
  pages={508--513},
  year={2017},
  publisher={American Association for the Advancement of Science}
}

@article{balduzzi2019open,
  title={Open-ended learning in symmetric zero-sum games},
  author={Balduzzi, David and Garnelo, Marta and Bachrach, Yoram and Czarnecki, Wojciech M and Perolat, Julien and Jaderberg, Max and Graepel, Thore},
  journal={arXiv preprint arXiv:1901.08106},
  year={2019}
}

@article{omidshafiei2020navigating,
  title={Navigating the Landscape of Games},
  author={Omidshafiei, Shayegan and Tuyls, Karl and Czarnecki, Wojciech M and Santos, Francisco C and Rowland, Mark and Connor, Jerome and Hennes, Daniel and Muller, Paul and Perolat, Julien and De Vylder, Bart and others},
  journal={arXiv preprint arXiv:2005.01642},
  year={2020}
}

@article{candogan2011flows,
  title={Flows and decompositions of games: Harmonic and potential games},
  author={Candogan, Ozan and Menache, Ishai and Ozdaglar, Asuman and Parrilo, Pablo A},
  journal={Mathematics of Operations Research},
  volume={36},
  number={3},
  pages={474--503},
  year={2011},
  publisher={INFORMS}
}

@inproceedings{heinrich2015fictitious,
  title={Fictitious self-play in extensive-form games},
  author={Heinrich, Johannes and Lanctot, Marc and Silver, David},
  booktitle={International Conference on Machine Learning},
  pages={805--813},
  year={2015} 
}

@inproceedings{lanctot2009monte,
  title={Monte Carlo sampling for regret minimization in extensive games},
  author={Lanctot, Marc and Waugh, Kevin and Zinkevich, Martin and Bowling, Michael},
  booktitle={Advances in neural information processing systems},
  pages={1078--1086},
  year={2009}
}

@article{koller1996finding,
  title={Finding mixed strategies with small supports in extensive form games},
  author={Koller, Daphne and Megiddo, Nimrod},
  journal={International Journal of Game Theory},
  volume={25},
  number={1},
  pages={73--92},
  year={1996},
  publisher={Springer}
}

@book{elias2012characteristics,
  title={Characteristics of games},
  author={Elias, George Skaff and Garfield, Richard and Gutschera, K Robert},
  year={2012},
  publisher={MIT Press}
}

@misc{heeb2012report,
  author={Heeb, Randall D}, 
  title={Report of Randall D. Heeb, PHD (United States of American against Lawrence Discristina)},
  howpublished={Case1:11-cr-00414 Document 77-1},
  month={7},
  year={2012}
}

@article{croson2008poker,
  title={Poker superstars: Skill or luck? Similarities between golf—thought to be a game of skill—and poker},
  author={Croson, Rachel and Fishman, Peter and Pope, Devin G},
  journal={Chance},
  volume={21},
  number={4},
  pages={25--28},
  year={2008},
  publisher={Taylor \& Francis}
}

@article{dedonno2008poker,
  title={Poker is a skill},
  author={DeDonno, Michael A and Detterman, Douglas K},
  journal={Gaming Law Review},
  volume={12},
  number={1},
  pages={31--36},
  year={2008},
  publisher={Mary Ann Liebert, Inc. 140 Huguenot Street, 3rd Floor New Rochelle, NY 10801~…}
}

@article{kelly2007poker,
  title={Poker and the law: is it a game of skill or chance and legally does it matter?},
  author={Kelly, Joseph M and Dhar, Zeeshan and Verbiest, Thibault},
  journal={Gaming law review},
  volume={11},
  number={3},
  pages={190--202},
  year={2007},
  publisher={Mary Ann Liebert, Inc. 2 Madison Avenue Larchmont, NY 10538 USA}
}

@article{van2015beyond,
  title={Beyond chance? The persistence of performance in online poker},
  author={van Loon, Rogier JD Potter and van den Assem, Martijn J and van Dolder, Dennie},
  journal={PLoS one},
  volume={10},
  number={3},
  year={2015},
  publisher={Public Library of Science}
}

\appendix

\section{Variance component formula derivation} \label{sec:decomp_derivation_value_based}

Recall that $\actind_z = \prod_{i \in \allplayers} \actind_z^i = \prod_{i \in \allplayers} \prod_{j=1}^{m^i(z)} I_{z,j}^i$ is the indicator that all actions along terminal history $z$ are selected, $Y = \sum_{z \in \mathcal{Z}} r(z) \, \actind_z$, and $u^i_{z,j}$ is the $j$th information state observed by player $i$ in terminal history $z$. Write $I_{z,k:}^i = \prod_{j=k}^{m^i(z)} I_{z,j}^i$, the indicator that player $i$ selects all actions in $z$ at and after $u^i_{z,k}$. Let $d(u)$ be the depth of $u$ in its trajectory; for example, if $u$ is the first observation of a player in their trajectory, $d(u)=1$. Define $W_u = \sum_{z \in \mathcal{Z}(u)} r(z) \, \eta^{-i}(z) \, \actind^i_{z,d(u):}$ and $\U^i_d = \{ u \in \U^i : d(u) = d\}$. Then 

\begin{align}
    V[E(Y|\mathbf{A}^i)] 
         = V\bigg[\sum_{z \in \mathcal{Z}} r(z) \, \eta^{-i}(z) \, \actind^i_z \bigg]
         = V\bigg[\sum_{u \in \U^i_1} \sum_{z \in \mathcal{Z}(u)}  r(z) \, \eta^{-i}(z) \, \actind^i_z \bigg] = V\bigg(\sum_{u \in \U^i_1} W_u \bigg). \label{eqn:var_component}
\end{align}
Note that histories $z \in \mathcal{Z}$ that contain no information states for player $i$ have $\actind^i_z \equiv 1$, so they are constant inside the conditional expectation, which is why the second and third expressions are equal.

By the perfect recall assumption, each information state $u$ can be uniquely identified with the sequence of information states and actions required to reach $u$. Furthermore, the behavioral strategy assumption gives that $A(u)$ is independent of $A(u')$ if $u \neq u' \in \U^i$. Therefore if $u^i_{z,j} \neq u^i_{z',j}$ for some $j$, then $I_{z,h}^i$ is independent of $I_{z',h'}^i$ for all $h,\, h' \in \{j,\dots, \min[m^i(z),m^i(z')]\}$. We conclude that $W_u$ is independent of $W_{u'}$ if $u \neq u'$. This allows us to split up \eqref{eqn:var_component}:
\begin{align}
    V[E(Y|\mathbf{A}^i)] 
         = V\bigg(\sum_{u \in \U^i_1} W_u \bigg)
         = \sum_{u \in \U^i_1} V(W_u) = \sum_{u \in \U^i_1} \bigg( V\{E[W_u | A(u)]\} + E\{V[W_u | A(u)]\} \bigg). \label{eqn:sum_of_wu}
\end{align}
The last equality holds by the law of total variance. To evaluate the components of \eqref{eqn:sum_of_wu}, write $W_{ua} = \sum_{z \in \mathcal{Z}(ua)} r(z) \, \eta^{-i}(z) \, \actind^i_{z,[d(u)+1]:}$ for each $a \in \A(u)$ so we have that 
\begin{align*}
    W_u 
    = \sum_{z \in \mathcal{Z}(u)} r(z) \, \eta^{-i}(z) \, \actind^i_{z,d(u):} 
    = \sum_{a \in \A(u)} \sum_{z \in \mathcal{Z}(ua)} r(z) \, \eta^{-i}(z) \, \actind^i_{z,d(u)} \actind^i_{z,[d(u)+1]:} 
    = \sum_{a \in \A(u)} W_{ua} \ind(A(u) = a).
\end{align*} 
Now evaluate each variance component. We begin with:
\begin{align*}
    V\{E[W_u| A(u)]\} &= V\bigg[E\bigg( \sum_{a \in \A(u)} W_{ua} \ind(A(u) = a)  \bigg| A(u) \bigg) \bigg] \\
    &= V\bigg[\sum_{a \in \A(u)} E(W_{ua}) \, \ind(A(u) = a)  \bigg] \\
    &= \sum_{a \in \A(u)} E^2(W_{ua}) \, \pi^i(a|u) - \bigg[\sum_{a \in \A(u)} E(W_{ua}) \, \pi^i(a|u) \bigg]^2.
\end{align*}
Write $r(u,a) = E\{r(Z) \ind[Z \in \mathcal{Z}(u,a)]\}$ and $r(u) = E\{r(Z) \ind[ Z \in \mathcal{Z}(u)]\}$. Then
\begin{align*}
    E(W_{ua}) &= \sum_{z \in \mathcal{Z}(ua)} r(z) \, \eta^{-i}(z) \, \textstyle\prod_{j=d(u)+1}^{m^i(z)} \pi(a_{z,j}^i | u_{z,j}^i)  \\
    &= \sum_{z \in \mathcal{Z}(ua)} r(z) \, \eta^{-i}(z) \, \textstyle\prod_{j=1}^{m^i(z)} \pi(a_{z,j}^i | u_{z,j}^i) \big[ \textstyle\prod_{j=1}^{d(u)} \pi(a_{z,j}^i | u_{z,j}^i) \big]^{-1} \\
    &= \sum_{z \in \mathcal{Z}(ua)} r(z) \, \eta(z) \, [ \eta^i(u) \pi(a|u) ]^{-1} \\
    &= r(u,a) \, [ \eta^i(u) \pi(a|u) ]^{-1}.
\end{align*}

Substituting these terms back into the expression for the variance component, we get that
\begin{align*}
    V\{E[W_u| A(u)]\} &= \sum_{a \in \A(u)} [r(u,a)]^2 \, [\eta^i(u)]^{-2} / \pi^i(a|u) - \bigg[[\eta^i(u)]^{-1} \sum_{a \in \A(u)} r(u,a) \bigg]^2 \\
    &= [\eta^{i}(u)]^{-2} \bigg( \sum_{a \in \A(u)} [r(u,a) ]^2 \, \pi^i(a ; u) - [r(u)]^2 \bigg).
\end{align*}
For the second variance component, we find:
\begin{align*}
    E\{V[W_u| A(u)]\} &= \sum_{a \in \A(u)} V[W_u| A(u)=a] \; P[A(u) = a] \\
    &= \sum_{a \in \A(u)} V\bigg(\sum_{z \in \mathcal{Z}(ua)} r(z) \, \eta^{-i}(z) \, \actind^i_{z,[d(u)+1]:} \bigg) \, \pi^i(a|u).
\end{align*}
Now take $Y_2 = \sum_{z \in \mathcal{Z}(ua)} r(z) \, \eta^{-i}(z) \, \actind^i_{z,[d(u)+1]:}$ and repeat the steps shown in \eqref{eqn:sum_of_wu} inductively to obtain that:
\begin{align*}
V[E(Y|\mathbf{A}^i)] &= \sum_{u \in \U^i}  \bigg( \sum_{a \in \A(u)} [r(u,a) ]^2 / \pi^i(a | u) - [r(u)]^2 \bigg) / \eta^{i}(u).
\end{align*}
Because $r(u,a) = q(u,a) \, \eta(u) \,  \pi^i(a|u)$ and $r(u) = v(u) \, \eta(u)$, this yields  \eqref{eqn:info_state_based_component}.

\section{Proof of consistency} \label{sec:consistency_proof}
Let $\mu(u) = \eta(u)  / \sum_{u \in \U^i} \eta(u)$. Then
\begin{align*}
     V[E(Y|\mathbf{A}^i)] &= \sum_{u \in \U^i}  \bigg( \sum_{a \in \A(u)} [q(u,a) ]^2 \, \pi^i(a | u) - [v(u)]^2 \bigg) \, \eta^{-i}(u) \, \eta(u) \\
     &= \bigg( \sum_{u \in \U^i} \eta(u)  \bigg)\sum_{u \in \U^i}  \bigg( \sum_{a \in \A(u)} [q(u,a) ]^2 \, \pi^i(a | u) - [v(u)]^2 \bigg) \, \eta^{-i}(u) \, \mu(u) \\
     &= \bigg( \sum_{u \in \U^i} \eta(u)  \bigg) E_{U \sim \mu} \bigg[ \bigg( \sum_{a \in \A(U)} [q(U,a) ]^2 \, \pi^i(a |U) - [v(U)]^2 \bigg) \, \eta^{-i}(U) \bigg]
\end{align*}

Note that 
\begin{align*}
    \sum_{u \in \U^i} \eta(u) = \sum_{u \in \U^i} \sum_{z \in \mathcal{Z}(u)} \eta(z) = \sum_{u \in \U^i} \sum_{z \in \mathcal{Z}} \eta(z) \ind(u \in z) =  \sum_{z \in \mathcal{Z}} \eta(z) \sum_{u \in \U^i} \ind(u \in z) = E[d^i(Z)],
\end{align*}
where $d^i(Z)$ is the length of the trajectory for player $i$ in terminal history $Z$. So, by the law of large numbers, $\nu^{-1} \sum_{k=1}^\nu d^i(Z_k) \overset{\text{a.s.}}{\rightarrow} \sum_{u \in \U^i} \eta(u)$ as $\nu \rightarrow \infty$. Consider the Markov Chain $\{U_t\}_{t \in \N}$ defined by the information states for player $i$ observed upon repeated independent playthroughs of the game and let $\phi(u) = \{ \sum_{a \in \A(u)} [q(u,a) ]^2 \, \pi^i(a |u) - [v(u)]^2 \} \, \eta^{-i}(u)$. Then $T^{-1} \sum_{t=1}^T \phi(U_t) \overset{\text{a.s.}}{\rightarrow} E_{U\sim \mu}[\phi(U)]$ as $T \rightarrow \infty$ by a Law of Large Numbers for Markov Chains since $\{U_t\}$ is irreducible and positive recurrent.

 Converting both these results to the notation of the original statement of the estimator, we have $\nu^{-1} \sum_{k=1}^\nu l_k \overset{\text{a.s.}}{\rightarrow} \sum_{u \in \U^i} \eta(u)$ and $\big( \sum_{k=1}^\nu l_k \big)^{-1} \sum_{k=1}^\nu \sum_{l=1}^{l_k} \phi(U_{k,l})  \overset{\text{a.s.}}{\rightarrow} E_{U\sim \mu}[\phi(U)]$ as $\nu \rightarrow \infty$. Therefore their product converges to the estimand, as desired.

\section{SkillRPS decomposition details} \label{sec:skillrps_details}
Recall that in SkillRPS, the outcome is $Y = (1-W)[\ind(S>0)-\ind(S<0)] + W(2Z-1)$, where $S=N_1-N_2+c \cdot \text{RPS}(A_1,A_2)$. In this case, a player's selection of $N_i$ is considered to indicate their skill level, and $\allactions^\text{c} = (W,Z)$ is the collection of all chance actions. Adapting the three-way decomposition equation \eqref{eqn:threeway_decomp} to SkillRPS yields
\begin{align*}
        V(Y) = \underbrace{V[E(Y|N_1,N_2)]}_{\text{skill}}+ \underbrace{E\{V[E(Y|W,Z,N_1,N_2)|N_1,N_2]\}}_{\text{chance}}+ \underbrace{E[V(Y|W,Z,N_1,N_2)]}_{\text{remaining variation}}.
\end{align*}
Under the assumption that $N_1,N_2 \iid \text{Uniform}(\{1,\dots,n\})$ and are independent of $A_1,A_2 \iid \text{Uniform}(\{ \text{Rock, Paper, Scissors}\})$, we can derive closed form expressions for each variance component.

Using routine probability manipulations, one can derive the following term for the variance in $Y$ explained by the ``skill'' of the players in the case that the coin flip didn't happen ($W=0$), for all $n \in \N$ and $c \in \N \cup \{0\}$. Begin by finding $E(Y|N_1=n_1,N_2=n_2,W=0)$ for arbitrary $n_1, n_2 \in \{1,\dots,n\}$, which is easy since the only remaining source of variation is $\text{RPS}(A_1,A_2) \sim \text{Uniform}(\{-1,0,1\})$. Next, treat this term as a discrete random variable depending on $N_1$ and $N_2$ and compute its variance. This yields:
\begin{align*}
    V[E(Y|N_1,N_2,W=0)] = 
    \begin{cases}
    1-\frac{1}{n} & \text{ if } c=0 \\
    1-\frac{1}{3n} + \frac{8c^2 + 2c - 16cn}{9n^2} & \text{ if } 0<c<n \\
    (1-\frac{1}{n})/9 & \text{ if } c \geq n.
    \end{cases}
\end{align*}
Call this term $\psi(n,c)$. From here one can find that:
\begin{align*}
    V[E(Y|N_1,N_2)] &= (1-\alpha)^2 \, \psi(n,c) \\
    E\{V[E(Y|W,Z,N_1,N_2)|N_1,N_2]\} &= \alpha + \alpha(1-\alpha)\,\psi(n,c).
\end{align*}
Finally, we also get that
\begin{align*}
    E[V(Y|W,Z,N_1,N_2)] = 
    \begin{cases}
    0 & \text{ if } c=0 \\
    (1-\alpha) \, [1-\frac{1}{n} + \frac{2c}{3n^2} - \psi(n,c)] & \text{ if } 0<c<n \\
    (1-\alpha) \, ( \frac{8}{9} - \frac{2}{9n}) & \text{ if } c \geq n.
    \end{cases}
\end{align*}

\end{document}